# Features of the Photon Pulse Detection Algorithm in the Quantum Key Distribution System

A.P. Pljonkin
Southern federal university
Russia, Taganrog
+79054592158
pljonkin@mail.ru

## ABSTRACT
A two-pass fiber-optic quantum key distribution system with phase-encoded photon states in synchronization mode has been investigated. The possibility of applying the analytical expressions for the calculation of the correct detection probability of the signal time window at synchronization has been proved. A modernized algorithm of photon pulse detection, taking into account the dead time of the single-photon avalanche photodiode was proposed. The method of engineering an optical pulse detection process during the synchronization in a quantum key distribution system has been offered.

## 1. INTRODUCTION
Classical cryptographic methods based on mathematics patterns and potentially can be broken (decrypted). Basic principles of quantum cryptography consist in absolute secrecy of transmitted messages and impossibility of unauthorized access to information an unauthorized person. Methods of quantum cryptography are based on quantum physics laws and realized in quantum key distribution systems (QKDS) [1-4].

Theoretical research of quantum key distribution systems was conducted by many laboratories [5-8], but to commercial realization brought several systems. Among the implemented systems are allocated a two-pass self-compensation fiber-optic system with phase coding states of photons, which are different stable operability. Such QKD systems consist of two stations, one of which is a transceiver and the second is coding station. Communication is organized by the fiber-optic communication line (FOCL) connecting the two stations, and control via personal computers. Before formation and distribution processes in QKDS, the transceiver and coding stations must be synchronized. Let us recall the principle of the so-called "plug&play" auto-compensating set-up, where the key is encoded in the phase between two pulses travelling from Bob to Alice and back (Figure 1 (a),(b),(c),(d)).

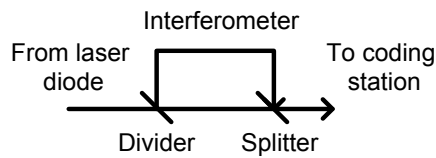

**Figure 1 (a) – Fiber-optic way in Bob QKDS station**

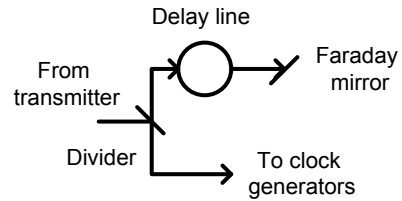

**Figure 1 (b) – Fiber-optic way in Alice QKDS station**

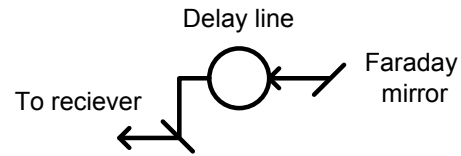

**Figure 1 (c) – Fiber-optic back way in the direction to Bob QKDS station**

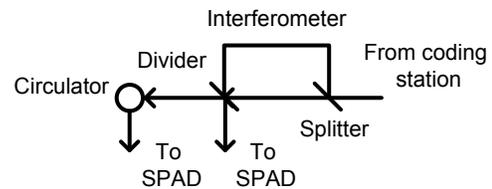

**Figure 1 (d) – Fiber-optic back way in Bob QKDS station**

The effective functioning of the QKDS possible only when synchronization is successful. The synchronization process is highly accurate measurement of the length of the propagation path of the optical radiation from transceiver station (Bob) to coding station (Alice) and back (Figure 2). In QKDS synchronization based on the moment of registration of the optical pulse receiving photodetectors formed on the basis of single-photon avalanche diodes (SPAD).

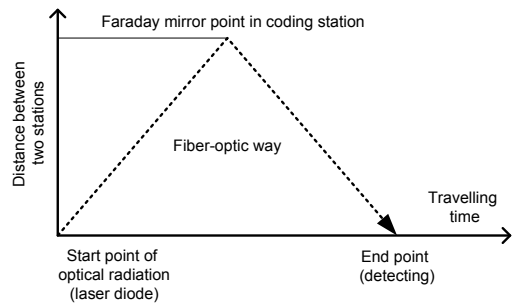

**Figure 2 – Optical pulse propagation way between stations**

The most suitable form of synchronization signal in QKDS is periodic sequence of optical pulses. Here the time markers are pulses themselves. The process of measuring the path length of the impulse propagation is dividing time frame on the time windows (subintervals). During the sync initialization, time frame equal to the impulse-repetition period of the optical impulse is divided into Nw time windows of fixed duration Tw (Figure 3). At synchronization (process of search of optical pulse) the duration of time interval (time window) is 2ns. Duration of optical pulse is about 1ns. Duration of impulse repetition period is about 1ms. Each time window is sampled N times, determining the amount of sampling. Time frame is the period of optical pulses. In every time window registered a conversion photon to electron or dark current pulses (DCP). The time window with the largest number of counts (registered photons and DCPs) taken as a "signal window" and the rest of the windows relate to "noise windows".

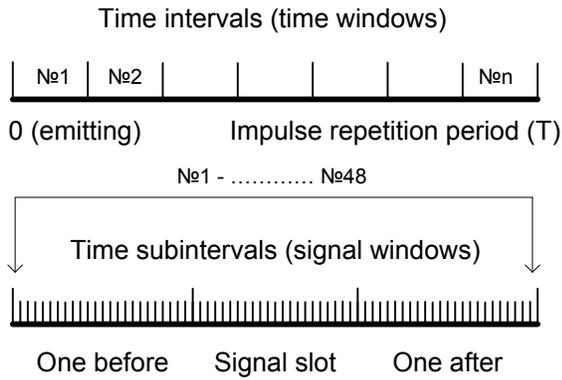

**Figure 3 – Optical pulse search algorithm at sync**

Figure 3 showed that in synchronization mode at first stage all time windows are analyzed. When the signal window(s) is detected, algorithm goes to second stage. At second stage of synchronization mode, algorithm is analyzed three time intervals (with signal window in center). Each time interval divided into 17 subintervals. Each subinterval is analyzed about 800 times. Results of detecting are formed.

In [9] the stand and experimental results of quantum-cryptographic network based on QKDS Clavis2 (idQuantique) are described. There is shown that synchronization in QKDS realized in multiphoton mode, where average number of photons per pulse is hundreds and thousands. This is consistent with results in [10], where showed that in synchronization mode the photodetectors works in linear mode. Implementation of the multiphoton mode synchronization potentially facilitates the organization of an attacker to unauthorized access to information.

This is determines a topicality of synchronization algorithms development with improved protection from unauthorized access.

In [11, 12] the new synchronization algorithm with improved protection from unauthorized access was proposed. The feature of algorithm is using a single photon pulse as synchronization signal, herewith the average photons number is not more 0,1 per pulse. For calculating the probability characteristics (the detection probability of signal time window) the analytical expressions (1), (2) were obtained. The (3) is simplified expression for (1).

$$P_D = \sum_{n_{w.N}=1}^{\infty} \frac{(\overline{n_{w.N}})^{n_{w.N}}}{n_{w.N}!} \cdot \exp[-\overline{n_{w.N}}] \cdot P_{d.N}\{n_{w.N}\}. \quad (1)$$

Where

$$P_{d.N}\{n_{w.N}\} = \left( \sum_{n_{d.N}=0}^{n_{w.N}-1} \frac{\overline{n_{d.N}}^{n_{d.N}}}{n_{d.N}!} \cdot \exp(-\overline{n_{d.N}}) \right)^{N_w - 1} \quad (2)$$

is probability of registered not more $(n_{w.N} - 1)$ DCP in all $(N_w - 1)$ noise windows during analysis on condition, that in signal time window during analysis (N) was registered $n_{w.N}$ photoelectrons and DCP.

$$\begin{aligned} P_D = \exp(-N_w \cdot \overline{n_{d.N}} + \overline{n_{d.N}}) \cdot (\overline{n_{w.N}} \cdot \exp(-\overline{n_{w.N}}) \\ + [1 - \exp(-\overline{n_{w.N}}) - \overline{n_{w.N}} \\ \cdot \exp(-\overline{n_{w.N}})] \cdot (1 + \overline{n_{d.N}})^{N_w - 1}). \end{aligned} \quad (3)$$

The simulation results showed that the difference between the calculation results on equations (1) - (3) does not exceed 0.02%. Also, it proves the possibility of using the expression (3) to calculate the probability of correct detection of the signal time window on condition, that $\overline{n_{w.N}} \ll 1$.

To assess the effectiveness of the proposed synchronization algorithm the simulations was performed. The simulation results proved the effectiveness of the proposed algorithm. Advantage of algorithm obvious at use (4) when the duration of the time window ($\tau_w$) exceeds the pulse duration ($\tau_s$) in 2 (4) times.

$$\tau_w = (2 \ldots 4) \cdot \tau_s. \quad (4)$$

At the same time, the research pointed out that as a photodetector is used a perfect single-photon device, which is able to record all incoming photons (photoelectrons). Such photodetectors do not need time to recover after photons or DCP registration. Features of single-photon avalanche diodes used in QKDS differ from the characteristics of the ideal single-photon photodetector. Firstly, SPAD registered only one (first) photon at analysis (work time in Geiger mode). Note that the time for analysis in the synchronization process refers to the duration of the time window. Secondly, in the case of photon registration for SPAD need some time to recover the working condition [13, 14].

**The purpose of research is to assess the impact of SPAD parameters on probability and time characteristics of the synchronization in QKDS.**

During synchronization, perhaps two cases of the location of the single-photon pulse in the time window: impulse completely located in the window or randomly located between two adjacent time windows. If a single-photon pulse is completely located at one time window, the correct detection is possible under two conditions. Firstly, in the signal time window during analysis must be registered at least one photon or DCP. Secondly, in the signal time window, the number of registered pulses must be strictly greater than the number of registered pulses in all other noise windows. The allocation of the single-photon pulse between two adjacent windows, correct detection is possible if the other conditions: registration at least one photon or DCP in one of the two windows containing the photon momentum; in the first time window signal number of registered pulses must be strictly greater than the number of registered pulses in the 2nd signal window and DCPs in all other noise windows; in the second signal window containing part of the photon momentum, the number of registered pulses is strictly greater than the number of registered pulses in the 1st signal window and DCPs in each of the noise windows; when equal number of accumulated pulses in two adjacent time windows a decision about registration of the photon pulse accepted to any of these windows, if the number of registered pulses in this window exceeds the number of registered pulses in the other windows.

## 2. THE TIME PARAMETERS

Let the known refractive index of the optical radiation into the optical fiber core ($n_{fiber}$ =1,49). Then the velocity of propagation of optical signals in a fiber optic link is

$$v_{fiber} = c_{opt}/n_{fiber} = 300000/1,49 = 201000 \text{ km/s},$$

The length of the fiber optic link between two QKDS stations $L_{FOL}$ it exceeds 100 km. Given the back-propagation of radiation in two-pass self-compensation fiber-optic system with phase coding states of photons to avoid imposition of counter pulses with $L_{FOL}$ =100 km the value of the optical pulse repetition period must be greater than $T_{s.min} = 2 \cdot L_{FOL}/v_{fiber} \approx$ 1 ms. Therefore, the maximum optical pulse repetition frequency at FOCL (quantum channel) extent of 100 km shall not exceed $f_{s.max} = 1/T_{s.min} \approx$ 1 kHz. The duration of the optical pulse is assumed to be $\tau_s = 1$ ns. For universal algorithm must convert events requiring counting during synchronization so that they are multiples of 2.

**Example design of synchronization subsystem QKDS.** Let $L_{FOL}$ = 100 km. For single mode optical fiber Corning® SMF-28e+ with wavelength 1550 nm $n_{fiber}$ =1,4670. Consequently, $v_{fiber} = c_{opt}/n_{fiber} \approx$ 205000 km/sec. Given the back-propagation of optical radiation in a two-pass system $T_{s.min} = 2L_{FOL}/v_{fiber} \approx$ 978 us. Therefore, the maximum frequency of the optical pulse should not exceed 1 kHz. Let $\tau_s = 1\ ns$ with $T_s = 1\ ms$ ($f_s = 1/T_s \approx$ 1 kHz). Based on the requirements (4) $\tau_w = 2\tau_s$=2 ns. The number of time slots is $N_w = T_s/\tau_w = 5 \cdot 10^5 = 500\ 000$. Therefore, the detection signal of the time window will enable 500 000 times to reduce the initial temporary uncertainty of the moment of the reception of synchronization signal.

The number of time slots is not a multiple of two, so the $T_s \geq$1 ms, $N_w$ =524 288=$2^{19}$. Therefore, $T_s = N_w \tau_w = 524\ 288 \cdot 2 = 1\ 048\ 576\ ns \approx 1,05\ ms$ ($f_s = 1/T_s \approx$ 954 Hz). Increasing the time frame does not exceed 5 %.

If the sample size is taken as N = 256 = $2^8$, the total time of the analysis of 256 temporary staff by using a single-photon ideal photodetector will be 256·1,05=268,8 ms. Using the SPAD id100-SMF20 with the frequency of occurrence of DCP less than 5 Hz (low noise version) will provide the average number of registered DCP for the sample volume in a noisy time window

$$\bar{n}_{d.N} = N \cdot \xi_d \cdot \tau_w = 256 \cdot 5 \cdot 2 \cdot 10^{-9} = 2,56 \cdot 10^{-6}.$$

When fiber optic line extent of 100 km the loss in the optical fiber will be 20 dB, i.e. a signal will weaken to 100 times. The average number of photoelectrons in a weak photon pulse, the receiving Bob station will be $\bar{n}_s = 0,001$. The average number of photoelectrons and DCP, is taken for the duration of photon pulses per sample in the signal time window, will be

$$\bar{n}_{w.N} = \bar{n}_{d.N} + N \cdot \bar{n}_s = 2,56 \cdot 10^{-6} + 256 \cdot 0,001 \cong 0,256.$$

Calculating the probability of correct detection of the signal of the time window for the approximate formula (3) gives $P_D = 7,95$ %. Given the weakening of the limit of the optical signal propagation in the reverse to the level of 0,001, increase the sample size to 1024. In view of this, we obtain the probability of detection of the expression (3) $P_D = 27,5\%$. At the same time, at a DCP frequency of 25 Hz, the average signal level of 0,01, the sample volume in 1024 will get the probability of correct detection by the formula (3) $P_D = 99,89\%$.

## 3. RECOVERY TIME (DEAD TIME)

During the calculations do not take into account the dead time of the photon registration $\tau_{dead}$, model whose value id100-SMF20 sample is 45 ns. This indicates that after the inspection time window will take time $\tau_w - \tau_{dead}$. The latter requires the introduction of the concept of the length of the module $\tau_m$, which may not be less than the dead time of the photon registration $\tau_{dead}$. If $\tau_m$=64 ns, whereas for the 1st cycle will be analyzed every 32 th ($2^6$) time window, i.e. inspection of the 1st, 33rd, 65th, etc. time window (Figure 4). After the last inspection $2^{19} : 2^6 = 2^{13}$ ($2^{19} : 2^6$) time window system must go to the second cycle analysis and has the 2nd, 34th, 66th, etc. time windows. Upon completion of the last 32-second cycle is complete examination of all time slots in the frame once. Note that the number of cycles is $N_c = \tau_m/\tau_w$.

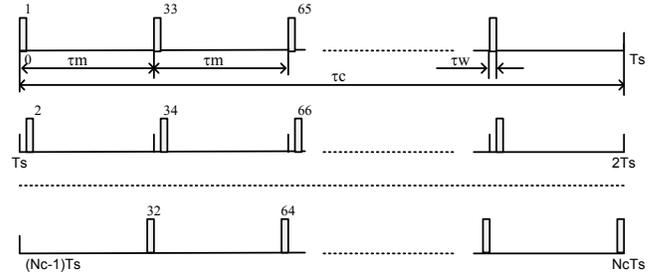

**Figure 4 - Search time window signal, taking into account the dead time of registration of the photon by SPAD**

The total analysis time with the required volume of sample 256, taking into account the dead time of registration of the photon by SAPD increases $N_c$ time (number of cycles).

At each sending of synchronization signal, the strob-signal activated SPAD, translating it into a single photon counting mode. This is calculated and fixed time delay of the gate signal. SPAD strobe-signal activates the duration of the time window. The expression (5) allows the calculation of the time delay detection in the detection signal of the time window.

$$Z_t = \frac{T_s}{4} \cdot (A_n - 1) + \tau_w \cdot (B_n - 1). \qquad (5)$$

where $A_n$ – serial number to activate the frame SPAD;

$T_s$ – impulse repetition period;

$B_n$ – frame sequence number.

Figure 5 shown a time diagram of detecting delay at synchronization mode. The impulse-repetition period value (Ts) is defined by the maximum extension of FOCL.

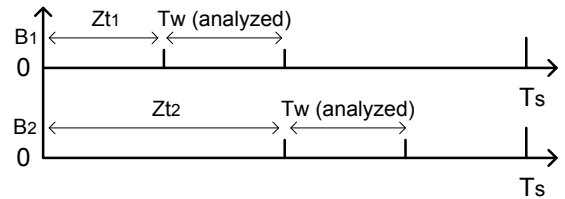

**Figure 5 – Time diagram of detection delay at sync**

In terms of probabilistic characteristics the position of the time window in which the signal pulse are detected, is of little importance.

## 4. CONCLUSIONS

The synchronization process of two pass self-compensation QKDS with phase-encoded photon states was described. A new algorithm for finding the signal time window with increased protection from unauthorized access was proposed. A feature of the algorithm is that as synchronization signal is used the attenuated photon pulse, where the average number of photons does not exceed 0,1. An analytical expression for engineering calculations of probability of correct detection of the signal window of time, reducing demands on computing resources was proposed. Search parameters of equipment and their impact on the detection signal of the time window during synchronization was analyzed. The method of designing the synchronization process taking into account the peculiarities of using a single-photon pulse as synchronization signal was proposed.

## 5. ACKNOWLEDGMENTS

The reported study was funded by RFBR according to the research project No.16-37-00003 mol_a.